\documentstyle[12pt]{article} 
\begin{document}%

\title{
Spin in a variable magnetic field:\\
the adiabatic approximation\\
{\em Spin dans un champ magn\'etique variable:\\
l'approximation adiabatique} }
\author{
 M. S. Marinov$^{a,b}$
and Eugene Strahov$^a$\\
(a) {\sl Physics Dept., Technion-Israel Institute of Technology}\\
{\it Haifa, 32000 Israel}\\
(b) {\sl LGCR, Universit\'e Pierre et Marie Curie - CNRS URA 7698}\\
{\it 4, Place Jussieu, 75005 Paris, France}
}
\date{(Submitted to ``{\em Comptes rendus Acad. Sci.Paris}".)}
\maketitle
   \begin{abstract}
The problem of spin precession in a time-dependent magnetic field
is considered in the adiabatic approximation where the field direction 
or the angular velocity of its rotation is changing slowly.
The precession angles are given by integrals in a way similar to the
semi-classical approximation for the Schr\"{o}dinger equation.


\vspace{3mm}

{\bf R\'esum\'e.} Le probl\`eme de la pr\'ecession du spin dans un champ 
magn\'etique variable est envisag\'e selon l'approximation 
adiabatique dans laquelle la direction du champ o\`u la vitesse de sa 
rotation varie lentement. Les angles de la pr\'ecession sont donn\'es par 
quadratures de mani\`ere similaire \`a l'approximation semi-classique de 
l'\'equation de Schr\"odinger.

{\bf Mots cl\'es:} 
prec\'ession magn\'etique du spin, approximation adiabatique
\end{abstract}

\subsubsection*{Version fran\c{c}aise abr\'eg\'ee}
Le probl\`eme de la pr\'ecession du spin dans un champ magn\'etique 
variable a \'et\'e r\'esolu explicitement pour deux cas particuliers 
importants (Landau and Lifshitz 1974, Sect. 114 ): 
(i) le champ magn\'etique ne change pas de direction, 
(ii) le champ de valeur constante tourne autour d'un axe 
donn\'e \`a une vitesse angulaire constante. En outre, des solution 
analytiques sont obtenues pour certaines variations du champ
(cf. Rosenfeld {\em et al.} 1996). La pr\'esente \'etude a pour but de 
pr\'esenter une solution approch\'ee de l'\'equation de Bloch (1) 
d\'ecrivant la pr\'ecession du spin. L'approximation est valable si la 
direction du champ ou de sa vitesse change adiabatiquement; 
le r\'esultat principal est \'enonc\'e par l'\'equation (16).

Nous consid\'erons l'\'equation de Bloch dans le cas o\`u la d\'ependance 
du champ magn\'etique ${\bf B}(t)$ de temps $t$ est arbitraire. La solution 
fondamentale est donn\'ee par une matrice unitaire $U(t,t_0)$ satisfaisant 
l'\'equation (3). Cette matrice peut \^etre \'ecrite en termes de deux 
fonctions complexes $(\xi,\eta)$, qui satisfont aux \'equations 
diff\'erentielles (5). Des \'equations de ce type et des solutions 
approch\'ees ont \'et\'e consider\'ees pr\'eced\'emment. Le probl\`eme 
peut \^etre r\'eduit \`a celui d'un oscillateur a fr\'equence variable et 
complexe, eq. (7). Il est \`a noter que le carr\'e de la fr\'equence 
d'oscillation dans l'eq. (7) est une constante r\'eele positive pour les 
cas simples, mais s'av\`ere \^etre g\'en\'eralement complexe. 
C'est pourquoi la m\'ethode traditionelle WKBJ ne peut \^etre appliqu\'ee 
directement.

Dans le but de trouver une bonne approximation, la matrice $U$ dans 
l'eq.(9) est d\'ecrite par une rotation d'un angle $2\alpha$ autour de la 
vecteur unitaire ${\bf n}=(\sin\gamma\cos\delta, \sin\gamma\sin\delta, 
\cos\gamma)$. Le probl\`eme consiste en recherche de la d\'ependance 
temporelle des trois angles $(\alpha,\gamma,\delta)$ \`a partir des 
\'equations diff\'erentielles (11)-(13).

L'hypoth\`ese cruciale, qui simplifie le probl\`eme, est que 
l'evolution de $\gamma$ est {\em tr\`es lente}. En d'autres termes, nous 
supposons que le vecteur $\bf n$ varie adiabatiquement. Les \'equations 
r\'esultantes sont integr\'ees immediatement par quadratures, eq. (16). Il 
est remarquable que pour les deux cas mentionn\'es dans le premier 
paragraphe, l'approximation est exacte, puisque soit (i) $B_\perp\equiv 0$, 
l'axe de rotation est toujours align\'e sur le champ, soit 
(ii) $\dot{\varphi}, B_z$ et $B_\perp$ sont fixes, de sorte que l'angle 
entre le champ et l'axe est preserv\'e, ainsi que l'angle $\gamma$. 

L'approximation edt effective sous deux conditions alternatives:
soit (i) la direction du champ $\bf B$, ou (ii) la direction du vecteur
$\bf (\dot{B}\times B)$, varient lentement autour d'un vecteur $\bf e$.
Dans les deux cas, le vecteur $\bf e$ indique l'axe $z$ pour le r\`epere 
appropri\'e et pour le choix des param\`etres de rotation dans 
l'\'equation (9).

La forme de la solution (16) obtenue ressemble \`a l'approximation 
semi-classique (ou eikonale) appliqu\'ee \`a la diffusion \`a haute 
energie, pour laquelle le potentiel de diffusion \'etait suppos\'e 
\^etre une fonction d'espace \`a variation lente.

$$^*{\scriptstyle *}^*$$
The problem of spin precession in variable magnetic field is fundamental 
for the theory of nuclear magnetic resonance(Abragam 1961). Its explicit 
solution is known for two important particular cases 
(Landau and Lifshitz 1974, Sect. 114 ): 
i) the magnetic field does not change its direction, 
ii) the field has a constant magnitude and is precessing around a given 
axis with a constant angular velocity. Besides, analytical solutions have 
been obtained for  a number of special field shapes (Rosen and Zener 1932, 
Bambini and Berman 1981, Silver {\em et al.} 1985, Rosenfeld {\em et al.} 
1996).

As soon as one deals with the Cauchy problem for two coupled linear
first-order differential equations, or equivalently, with a second-order 
equation, an adiabatic approximation may be employed, similarly 
to the standard semi-classical approach to the Schr\"{o}dinger equation.
An adiabatic approximation for the Bloch equation was
the purpose of the present work, and the result is presented in 
Eq.(16). The key to the desired approximation is a proper 
group-theoretical parametrization in Eqs.(9-10), which enables one to  
handle complexities specific to the problem.

The spinor wave function of a neutral spin-$\frac{1}{2}$ particle 
satisfies the Bloch equation,
   \begin{equation}  
id\psi/dt=({\bf B\cdot\sigma})\psi,
   \end{equation}
where ${\bf B}=(B_x,B_y,B_z)\equiv\mu{\cal B}(t)$, ${\cal B}(t)$ is 
the variable magnetic field vector, $\mu$ is the particle magnetic 
moment, and $\sigma$ are the Pauli matrices. The fundamental solution of 
Eq. (1) is given by a unitary $2\times 2$ matrix $U$,
   \begin{eqnarray}  
\psi(t)=U(t,t_0)\psi(t_0),\\
i\partial U/\partial t=({\bf B\cdot\sigma})U,\;\;\;
U(t_0,t_0)=I.
   \end{eqnarray}
(Here $I$ is the unit matrix.) Two complex functions $(\xi,\eta)$  
are used to represent $U$,
   \begin{equation}  
U=\left(\begin{array}{ll}
\xi e^{-i\beta} & \bar{\eta} e^{-i\beta}\\
-\eta e^{i\beta} & \bar{\xi} e^{i\beta}
\end{array} \right);\;\;\;\beta=\int^t_{t_0}B_z(\tau)d\tau.
   \end{equation}
In view of Eq. (3), these functions satisfy the following equations,
   \begin{equation}  
i\dot{\xi}=-b\eta,\;\;\;i\dot{\eta}=-\bar{b}\xi,
\;\;\;|\xi|^2+|\eta|^2=1.
   \end{equation}
where
   \begin{equation}  
b\equiv (B_x-iB_y)e^{2i\beta}=B_\perp e^{2i\beta-2i\varphi},
\;\;\;B_\perp^2={\bf B}^2-B_z^2,\;\;\;\tan 2\varphi=B_y/B_x.
\end{equation}
Equations of this type and various approximations were considered
previously (e.g. by Popov 1962, Marinov and Popov 1979, Rosenfeld 
{\em et al.} 1996). The problem may be further reduced to that for an 
oscillator with a variable (complex) frequency,
   \begin{equation}  
\ddot{f}+\Omega^2(t)f=0,\;\;\;f\equiv b^{-1/2}\xi,
  \end{equation}
where
   \begin{eqnarray}  
\Omega^2=B_\perp^2+\omega^2+i\dot{\omega},\;\;\;
\nonumber\\
\omega\equiv -i\dot{b}/2b=B_z-\dot{\varphi}-
\frac{i}{4}\frac{d}{dt}\log B_\perp^2,
   \end{eqnarray}
i.e. $2\dot{\varphi}=(B_x\dot{B}_y-\dot{B}_xB_y)/B_\perp^2$ 
is the angular velocity of the field rotation around the $z$-axis.
Note that $\Omega^2$ is a real positive constant for the cases presented 
in the textbook, but is complex in general. Therefore the standard WKBJ
method cannot be applied to Eq. (7) directly.

A consistent adiabatic approximation for the problem concerned was
proposed by Bender and Papanicolaou 1988 and further developed by
Papanicolaou 1988. It was assumed that the field time dependence was slow,
i.e. the specific variation time $T$ was large, and an expansion in
powers of $1/T$ was constructed. An alternative approach, based upon the
Magnus expansion, was presented by Klarsfeld and Oteo 1992. Unlike the
above mentioned (and a number of other) works, we do not assume a slow
time dependence of the field. In particular, the result given below is
{\em exact} for the field rotation with an arbitrarily high (but stable)
angular velocity. Actually, we employ geometrical arguments, which
were previously developed for the adiabatic approximation
by Nakagawa 1987. With all that, to the best of our knowledge,
the result given in Eqs. (9), (16) was never published before.

The complex equations (5) are equivalent to real equations for angles
which appear when $U$ is treated as the spinor representation of the 
rotation group. Let us write $U$ as a product of two matrices 
representing rotations,
   \begin{equation}  
U\equiv R_z(\alpha_1)R_n(\alpha),\;\;\;
R_n=I\cos\alpha-i({\bf n\cdot\sigma})\sin\alpha.
  \end{equation}
Here matrix $R_z(\alpha_1)$ has the diagonal elements 
$(e^{i\alpha_1},e^{-i\alpha_1})$ only and represents the rotation around 
the $z$-axis, and $R_n(\alpha)$ represents a rotation by an angle 
$2\alpha$ around the direction of a unit vector
${\bf n}=(\sin\gamma\cos\delta,\sin\gamma\sin\delta,\cos\gamma)$.
The factor containing $\alpha_1(t)$ transforms the system to a rotating
frame. The time dependence of $\alpha_1$ is chosen in such a way as to
make the variation of $\bf n$ slow. Setting
   \begin{equation}  
\alpha_1=-\varphi+\delta/2,
   \end{equation}
we get the following differential equations for the angles
$(\alpha,\gamma,\delta)$,
   \begin{eqnarray}  
\dot{\alpha}\sin\gamma=\nu\sin(2\gamma-\gamma_0),\\
\dot{\gamma}\tan\alpha=2\nu\sin(\gamma_0-\gamma),  \\
\dot{\delta}\sin\gamma=2\nu\sin(\gamma_0-\gamma),
   \end{eqnarray}
where the new notations are
   \begin{equation}  
\nu(t)=\sqrt{\dot{\varphi}^2-2\dot{\varphi}B_z+{\bf B}^2},\;\;\;
\cot\gamma_0(t)=(\dot{\varphi}-B_z)/B_\perp.
   \end{equation}
The initial conditions corresponding to (3) are
   \begin{equation}  
\alpha(t_0)=0,\;\;\;
\gamma(t_0)=\gamma_0(t_0),\;\;\;
\delta(t_0)=\varphi(t_0).
   \end{equation}
The starting precession velocity of $\bf n$, as given by Eqs. (12),(13),
is vanishing, and the crucial assumption is that it is small for all $t$.
In the other words, we assume that the vector $\bf n$ is moving
adiabatically. In contrast to the standard adiabatic approximation,
$\bf n$ does not coincide with the field direction if its change is
rapid. As seen from Eq. (14), $\bf n$ takes a position intermediate
between $\bf B$ and its instantaneous rotation axis 
$\bf(\dot{B}\times B)$. The approximate solution of (11)-(13) is
   \begin{equation}  
\alpha(t,t_0)=\int^t_{t_0}\nu(\tau)d\tau,\;\;\;
\gamma(t,t_0)\equiv\gamma_0(t),\;\;\;
\delta(t,t_0)\equiv\varphi(t_0).
\end{equation}
For two cases mentioned in the first paragraph the approximation is exact,
since either i) $B_\perp=0$, so $\gamma_0\equiv 0$, or
ii) $\dot{\varphi}, B_z$, and $B_\perp$ are fixed, so $\gamma_0$ is a
constant too.

In order to estimate the accuracy of the approximation, let us assume 
that $\dot{\gamma}_0$ is small and $\gamma\approx\gamma_0$. Respectively, 
equation (12) is linearized and solved explicitly
   \begin{eqnarray}  
\dot{\gamma}\approx 2(\gamma_0-\gamma)\dot{\alpha}\cot\alpha,\nonumber\\
\gamma(t,t_0)\approx\gamma_0(t)+\left[\int^t_{t_0}
\dot{\gamma_0}(\tau)\sin^2\alpha(\tau,t_0)d\tau\right]/
\sin\alpha^2(t,t_0),
   \end{eqnarray}
where $\alpha$ is given by (16).
The approximation is effective under two alternative conditions: either
(i) the field direction, or (ii) the direction of the vector 
$\bf (\dot{B}\times B)$, is changed slowly around a fixed vector $\bf e$.
In both cases, the vector $\bf e$ indicates the $z$ axis for
the appropriate coordinate system and the choice of the rotation 
parameters in Eq. (9). A remarkable case is that of a fastly rotating field,
where $\dot{\varphi}\gg B_\perp$, and $|\cot\gamma_0|\gg 1$. As a result,
the deviation of $\bf n$ from the $z$ direction is small, even if the field
rotation is not stationary. 
The obtained form of the solution (16) resembles the semi-classical and 
the eikonal approximation to the high-energy scattering (Landau and Lifshitz,
1974, Sect. 131), since it is valid if $\dot{\nu}\ll\nu^2$.

The validity of our approximation was checked by comparing it
with the exact result for a field where the analytical solution was
obtained by Rosen and Zener 1932. That field has a constant $x$-component
and a pulse in the $y$-direction,
    \begin{equation}  
{\bf B}(t)=\frac{1}{T}\left(\beta_0,\frac{\zeta}
{\cosh(t/T)},0\right),
\end{equation}
where $\beta_0$ and $\zeta$ are constant dimensionless parameters.
Assuming that the initial state was polarized to the $x$-axis, the
spin-flip probability $W$ was calculated in the time asymptotics, i.e.
$t,|t_0|\gg T$. In the limit of $\beta_0\rightarrow\infty$, corresponding
to a large $T$ for a fixed background field, the exact and the approximate
solutions coincide,
\begin{equation}
W=\zeta^4/\beta_0^2+O(\beta_0^{-3}).
\end{equation}
Moreover, the numerical calculations showed that the relative error of
our approximation in $W$ is less than 1\% for all values of 
$\beta_0\ge 1.5$ and $\zeta\ge 1$.

{\bf Acknowledgements.} We thank Sh. Panfil for interest and a helpful 
discussion. The support from the Technion V. P. R. Fund is gratefully
acknowledged.

\end{document}